\documentclass[12pt]{article}

\renewcommand{\thefootnote}{\fnsymbol{footnote}}
\usepackage{amsbsy,amssymb,latexsym,amsfonts,amsmath}
\usepackage{mathrsfs}
\usepackage{graphicx}
\usepackage{bm}
\usepackage{here}
\numberwithin{equation}{section}
\newcommand{\bel}[1]{\begin{equation}\label{#1}}                     
\newcommand{\bal}[1]{\begin{eqnarray}\label{#1}}                     
\newcommand{\be}{\begin{equation}}
\newcommand{\ee}{\end{equation}}

\newcommand{\ex}{\mathrm{e}}
\newcommand{\de}{\mathrm{d}}

\newcommand{\dis}{\displaystyle}

\newcommand{\qq}{\qquad}

\newcommand{\bea}{\begin{equation}}
\newcommand{\eea}{\end{equation}}

\begin{document}
%%%%%%%%%%%%%%%%%%%%%%%%%%%%%%%%%%%%%%%%%%%%%%%%%%%%%%%%%%%%%%%%%%%%%%%%%%%%%%%%
%%%%%%%%%%%%%%%%%%%%%%%%%%%%%%%%%%%%%%%%%%%%%%%%%%%%%%%%%%%%%%%%%%%%%%%%%%%%%%%%%%%%%%%%%%
%
% title page
%
%%%%%%%%%%%%%%%%%%%%%%%%%%%%%%%%%%%%%%%%%%%%%%%%%%%%%%%%%%%%%%%%%%%%%%%%%%%%%%%%%%%%%%%%%
\begin{titlepage}
%%%%%%%%%%%%%%%%%%%% preprint # %%%%%%%%%%%%%%%%%%
\begin{flushright}
\normalsize
%\filename
~~~~
OCU-PHYS 440\\
February, 2016 \\
\end{flushright}
%%%%%%%%%%%%%%%%%%%%%%%%%%%%%%%%%%%%%%%%%%%%%%%%%%

\vspace{15pt}

%%%%%%%%%%%%%%%%%%%% title %%%%%%%%%%%%%%%%%%%%%%%
\begin{center}
{\LARGE $q$-Vertex Operator from 5D Nekrasov Function}
\end{center}
%%%%%%%%%%%%%%%%%%%%%%%%%%%%%%%%%%%%%%%%%%%%%%%%%%

\vspace{23pt}

%%%%%%%%%%%%%%%%%%% authors %%%%%%%%%%%%%%%%%%%%%%
\begin{center}
{ H. Itoyama$^{a, b}$\footnote{e-mail: itoyama@sci.osaka-cu.ac.jp},
T. Oota$^b$\footnote{e-mail: toota@sci.osaka-cu.ac.jp}
  and  R. Yoshioka$^b$\footnote{e-mail: yoshioka@sci.osaka-cu.ac.jp}  }\\
%%%%%%%%%%%%%%%%%%%%%%%%%%%%%%%%%%%%%%%%%%%%%%%%%%
%
\vspace{18pt}
%
%%%%%%%%%%%%%%%%%%% affiliation %%%%%%%%%%%%%%%%%%%

$^a$\it Department of Mathematics and Physics, Graduate School of Science\\
Osaka City University\\
\vspace{5pt}

$^b$\it Osaka City University Advanced Mathematical Institute (OCAMI)

\vspace{5pt}

3-3-138, Sugimoto, Sumiyoshi-ku, Osaka, 558-8585, Japan \\

\end{center}
%%%%%%%%%%%%%%%%%%%%%%%%%%%%%%%%%%%%%%%%%%%%%%%%%%%
%
\vspace{20pt}
\begin{center}
Abstract\\
\end{center}
The five dimensional AGT correspondence implies the connection between
the $q$-deformed Virasoro block and the 5d Nekrasov partition function.
In this paper, we determine a $q$-deformation of the four-point block in the Coulomb gas representation
from the 5d Nekrasov function, and obtain an expression of the $q$-deformed vertex operator.
If we use only one kind of the $q$-vertex operators,
one of the insertion points of them must be modified in order to hold the 2d/5d correspondence.
%%%%%%%%%%%%%%%%%%%% abstract %%%%%%%%%%%%%%%%%%%%%

%%%%%%%%%%%%%%%%%%%%%%%%%%%%%%%%%%%%%%%%%%%%%%%%%%%

\vfill

\end{titlepage}

%%%%%%%%%%%%%%%%%%%%
\renewcommand{\thefootnote}{\arabic{footnote}}
\setcounter{footnote}{0}
%%%%%%%%%%%%%%%%%%%%

%%%%%%%%%%%%%%%%%%%%%%%%%%%%%%%%%%%%%%%%%%%%%%%%%%%%%%%%%%%%%%%%%%%%%%%%%%%%%%%%
%%%%%%%%%%%%%%%%%%%%%%%%%%%%%%%%%%%%%%%%%%%%%%%%%%%%%%%%%%%%%%%%%%%%%%%%%%%%%%%%
\section{Introduction}

Twenty years ago,  a $q$-deformation of the Virasoro algebra ($q$-Virasoro algebra) was introduced in 
\cite{LP94,FR95,SKAO}\footnote{
A $q$-deformation of $W$ algebra of type $A_n$ is treated in \cite{FF95,AKOS95}.
}. It is closely related to the one dimensional XYZ Heisenberg chain model
or to a two-dimensional solvable lattice model (the Andrews-Baxter-Forrester model).
One motivation for the $q$-deformation is to study thermodynamic limit of these models
using the representation theory of this deformed algebra.

In ordinary minimal conformal field theories (CFT), the singular vectors have connection
with the Jack symmetric functions indexed by a rectangular partition \cite{MY95}.
There is a generalization of the Jack functions, called Macdonald symmetric functions \cite{mac}.
The guiding principle for deformation in \cite{SKAO} is such that the singular vectors
of the deformed algebra
are expressed in terms of the Macdonald functions.
The defining relation of the $q$-Virasoro algebra and the screening currents
are well established in \cite{SKAO}.

On the other hand,  the representations of the $q$-Virasoro algebra is still not well understood.
In particular,  we have no proper definition of
$q$-deformed primary fields (vertex operators).
The vertex operators or intertwining operators of the $q$-Virasoro algebra
for the minimal cases 
are considered in \cite{AKMOS9604,kad,JS97}.
These $q$-vertex operators have ``good'' commutation relations
with the $q$-Virasoro generators.
But general criterion for goodness is not known.

Recently, the $q$-Virasoro algebra 
has collected renewed interests
due to the $q$-deformed/lifted version (or K-theoretic five dimensional version)
of the (W)AGT conjecture. The (W)AGT relation \cite{AGT,Wyl0907}
implies the two dimensional correlation functions (conformal blocks) of the Virasoro/W algebra 
are identical to the four dimensional Nekrasov partition functions.
For this 2d/4d correspondence, see for example \cite{MMM0907,DV,MMM0909b,IMO,MMS0911,FL0912,
IO5,IOYone,AFLT1012,Wy1109,EPSS1110,ZM1110,BBFLT1111,
KMZ1207,tan1301,MS1307,
MMZ1312,MRZ1405,IY1507,NZ1511}.
The deformed/lifted (W)AGT suggests that the $q$-deformed Virasoro/W blocks in two dimensional theories
are identified with the Nekrasov partition function of the five dimensional gauge theories
\cite{AY0910,AY1004}.
See also  \cite{MMSS1105,NPP1303,Tan1309,NPPT1312,ohk1404,zen1412,AFO1512,
KP1512,MZ1510} for the 5d AGT.
More general 2d/6d correspondence is also discussed in \cite{Tan1309,IKY1511,nie1511,MMZ1512}.

Assuming the $q$-deformed/lifted version of (W)AGT conjecture, one can fix
the form of the $q$-vertex operator. The conformal blocks have a Coulomb gas
representation (the Dotsenko-Fateev integral representation). 
A simple recipe for the $q$-deformation of the conformal block in the Coulomb gas
representation is proposed in \cite{MMSS1105} without constructing the $q$-operators.
We consider the problem of operator realization of the deformed block in order to determine
the $q$-vertex operator. 

In \cite{IO5}, it is shown that the Dotsenko-Fateev representation of the
four-point conformal block is related to 
multiple integrals with the Selberg measure.
The Kadell's formula
 gives the average of the Jack polynomials with respect to this measure.
 Using the formula,
we compared the conformal block with the 4d Nekrasov function and found agreement.

In this paper, we consider a straightforward $q$-deformation of \cite{IO5}.
Kaneko obtained a $q$-deformed version of the Kadell's formula \cite{kan96}.
This formula gives the average of the Macdonald polynomials
with respect to the $q$-deformed Selberg measure. By adopting it
as a calculational tool, and starting from the 5d $SU(2)$ Nekrasov function,
we obtained the $q$-deformed (Jackson) integral representation of the four-point block.
Using a free field representation of the $q$-screening charges, we have determined
an explicit form of the $q$-vertex operator. This is one of our main results.

 If we use only one kind of the $q$-vertex operators,
 it turns out that one of the insertion points of them must be modified
 in order to match the $q$-block with the Nekrasov function.
 We have no simple explanation on this modification.

This paper is organized as follows.
In the next section, after reviewing the $q$-Virasoro algebra and its screening charges,
we give explicitly the vertex operator for the $q$-Virasoro algebra
 such that the 2d/5d correspondence is established. 
The $q$-deformed version of the Coulomb gas representation of the four-point block is constructed. 
In Section 3, a brief review of the 5d Nekrasov partition function is given. 
In Section 4, we perform the $\Lambda_0$-expansion of the $q$-Virasoro block.  
The parameter dictionary of 2d/5d correspondence is presented. 
We display the explicit form of the first order term in $\Lambda_0$ expansion. 
In Section 5, we carry out the comparison with the 5d Nekrasov partition function. 
In this process, we obtain non-trivial relations among some quantities.    
In Appendix A,  the two-point correlation function for the $q$-Virasoro vertex operator 
and the $q$-screening current are given.
In Appendix B, the $q$-Selberg integral and Kaneko's formula are presented.

%%%%%%%%%%%%%%%%%%%%%%%%%%%%%%%%%%%%%%%%%%%%%%%%%%%%%%%%%%%%%%%%%%%%%%%%%%%%%%%%
%%%%%%%%%%%%%%%%%%%%%%%%%%%%%%%%%%%%%%%%%%%%%%%%%%%%%%%%%%%%%%%%%%%%%%%%%%%%%%%%
\section{$q$-deformed conformal block}

In this section, we first explain our convention for the $q$-deformed Virasoro algebra
and its screening charges. 
Then, we propose a $q$-deformed vertex operator. 
We also introduce a ``modified'' four-point block $\mathcal{B}(\Lambda_0)$.

%%%%%%%%%%%%%%%%%%%%%%%
\subsection{$q$-Virasoro algebra}

Let us consider a $q$-Heisenberg algebra with generators
\be
\alpha_n\ \ (n \in \mathbb{Z}), \qq Q
\ee
satisfying the following defining relations:
\be
[ \alpha_n, \alpha_m ] = - \frac{1}{n} \frac{(1-q^n)(1-t^{-n})}{(1+p^n)}
\delta_{n+m,0}, \qq (n \neq 0),
\ee
\be
[ \alpha_n, Q ] = \delta_{n,0}.
\ee
Here $p=q/t$.  The $q$-deformed Virasoro algebra can be realized by this algebra \cite{SKAO}
\be
\mathcal{T}(z) = : \exp\left( \sum_{n \neq 0} \alpha_n z^{-n} \right): p^{1/2} q^{\sqrt{\beta} \alpha_0}
+ : \exp\left( - \sum_{n \neq 0} \alpha_n (pz)^{-n} \right): p^{-1/2} q^{-\sqrt{\beta} \alpha_0}.
\ee
Here $\beta$ is defined by the relation $t=q^{\beta}$.

This operator satisfies the defining relation of the $q$-Virasoro algebra:
\be
f(z'/z) \mathcal{T}(z) \mathcal{T}(z') - f(z/z') \mathcal{T}(z') \mathcal{T}(z) = \frac{(1-q)(1-t^{-1})}{(1-p)}
\Bigl[ \delta(pz/z') - \delta(p^{-1} z/z') \Bigr],
\ee
where
\be
f(z) = \exp\left( \sum_{n=1}^{\infty} \frac{1}{n} \frac{(1-q^n)(1-t^{-n})}{(1+p^n)} z^n \right), \qq
\delta(z) = \sum_{n \in \mathbb{Z}} z^n.
\ee

In the following part, we assume that $|q|<1$.

%%%%%%%%%%%%%%%%%%%%%%%%%%%%%%
\subsection{Screening charges}

A screening current for the $q$-Virasoro algebra is defined by
\bel{qSCp}
S_+(z) = : \ex^{\tilde{\varphi}^{(+)}(z)}:,
\ee
where
\be
\tilde{\varphi}^{(+)}(z)
= \sqrt{\beta} Q + 2 \sqrt{\beta} \alpha_0 \log z
+ \sum_{n \neq 0} \frac{1+p^{-n}}{1-q^n} \alpha_n z^{-n}.
\ee
This screening current $S_+$ commutes with the $q$-Virasoro generators up to a total $q$-derivative:
\be
[ T(z), S_+(z') ]
= - (1-q)(1-t^{-1}) \frac{\de_q}{\de_q z'} \Bigl[ \delta(z'/z) p^{-1/2} z' A_+(z') \Bigr],
\ee
where
\be
A_+(z) = : \exp\left( \sum_{n \neq 0} \frac{(1+t^n)}{(1-q^n)} \alpha_n z^{-n} \right):
\ex^{\sqrt{\beta} Q} q^{-\sqrt{\beta} \alpha_0} z^{2\sqrt{\beta} \alpha_0}.
\ee
The $q$-derivative is given by
\be
\frac{\de_q  f(z)}{\de_q z} = \frac{f(z) - f(qz)}{z-qz}.
\ee
There is another screening current $S_-(z)$ which commutes with the $q$-Virasoro generators
up to a total $t$-derivative. But we will not use $S_-(z)$ in this paper, hence we do not introduce
it.
 
Using two different ``integration ranges'',
two screening charges $Q_+$, $Q_+'$ are defined by 
\be
Q_+:=\int_0^{\Lambda_0} \de_q z\, S_+(z),
\qq
Q_+':= \int_1^{\infty} \de_q z \, S_+(z)
\equiv \int_0^{1} \frac{\de_q y}{y^2} \, S_+(1/y).
\ee
Here we use the Jackson integral
\be
\int_0^a \de_q z\, f(z) = (1-q) \sum_{n=0}^{\infty} f(a q^n) a q^n.
\ee

%%%%%%%%%%%%%%%%%%%%%%%%%%%%%%%%%%%
\subsection{Vertex operators for the $q$-Virasoro algebra}

Vertex operators for the $q$-Virasoro algebra are considered in \cite{AKMOS9604,kad,JS97,AY1004}.
In \cite{AKMOS9604,kad,JS97}, a $q$-deformation of primary operators 
$V_{r,s}(z)=:\ex^{(1/2) \alpha_{r,s} \phi(z)}:$ in the
minimal CFT model is introduced. Here $\alpha_{r,s} = (1-r) \sqrt{\beta} - (1-s)/\sqrt{\beta}$ with
rational $\beta$. The normalization of chiral boson is chosen as $\langle \phi(z) \phi(z') \rangle = 2 \log(z-z')$.
But we are not interested in these types of $q$-vertex operators.

The $q$-vertex operator determined from the 5d Nekrasov function is the following.
For a complex parameter $u$, 
let us define a vertex operator of the $q$-deformed Virasoro algebra by
\bel{qVop}
V_{u}(z):= : \ex^{\Phi_u(z)}:,
\ee
where
\be
\Phi_u(z):= \frac{u}{\sqrt{\beta}}
\left( \frac{1}{2} Q + \alpha_0 \log z \right)
+ \sum_{n\neq 0} \frac{(q^{nu}-1)}{(1-q^n)(1-t^{-n})} \alpha_n z^{-n}.
\ee
This is essentially equivalent to the vertex operator $V^1_U(z)$ in \cite{AY1004}\footnote{By identification of the fundamental bosons $h^1_n$ in \cite{AY1004}
with our $\alpha_n$, we have $V_{q^{-u}}^1(z) = V_{u}(q^{u/2}z) q^{u^2 \alpha_0/(2\sqrt{\beta})}$.}\ 
with $U=q^{-u}$.

In the $q \rightarrow 1$ limit with keeping $\beta= \log t/\log q$ fixed, 
this vertex operator becomes a free field representation
of the Virasoro primary operator with scaling dimension $\Delta = u (u-2\beta+2)/(4\beta)$.

%%%%%%%%%%%%%%%%%%%%%%%%%%%%%%%%%%%
\subsection{$q$-deformed Coulomb gas representation}
\label{qDCGR}

A $q$-deformed version of the Coulomb gas representation
of the four-point block is defined by
\be
\langle  V_{u_1}(0) V_{u_2}(\Lambda_0)
V_{u_3}(1) V_{u_4}(\infty) (Q_+)^{N_+}
(Q_+')^{N_-} \rangle.
\ee
But we will consider the following ``modified'' four-point block:
\bel{4ptCB}
\mathcal{B}(\Lambda_0)
= \langle V_{u_1}(0) V_{u_2}(\Lambda_0)
V_{u_3}(q^{u_3+1}) V_{u_4}(\infty) (Q_+)^{N_+}
(Q_+')^{N_-} \rangle.
\ee
The four vertex operators are inserted at $z_1=0$, $z_2=\Lambda_0$, $z_3=q^{u_3+1}$ and $z_4=\infty$.
Note that $z_3=q^{u_3+1}$, instead of $z_3=1$.
We have no simple explanation on the position of the third vertex operator.
With this choice of $z_3$, the modified four-point conformal block \eqref{4ptCB} coincides with the five dimensional Nekrasov partition function of the $SU(2)$ gauge theory with $N_f=4$.

In  \eqref{4ptCB}, the parameters $u_i$ should obey
the ``momentum conservation condition'':
\bel{MCC}
u_1 + u_2 + u_3 + u_4 + 2 \beta( N_+ + N_-) + 2 (1-\beta) = 0.
\ee
The parameters $u_i$ used here is related to the 
parameters $\alpha_i$ in \cite{IO5} as
$u_i = \sqrt{\beta} \alpha_i.$

For simplicity, we assume that 
\be
0 < | \Lambda_0| < |q^{u_3+1}| < 1.
\ee
Using the two-point correlators in Appendix \ref{TPCF}, we have
\be
\begin{split}
\mathcal{B}(\Lambda_0)
&= \mathcal{V}_0(\Lambda_0) \, (N_+)! \, (N_-)! \int_{\mathcal{C}_{N_+}([0,\Lambda_0])}
\de_q^{N_+} z \,\int_{\mathcal{C}'_{N_-}([1,\infty])} \de_q^{N_-} z' \cr
& \times \prod_{i=1}^{N_+} z_i^{u_1}
\frac{(q^{-u_3} z_i; q)_{\infty} (q z_i/\Lambda_0; q)_{\infty}}
{( z_i; q)_{\infty} (q^{u_2+1} z_i/\Lambda_0; q)_{\infty}} 
\prod_{j=1}^{N_-} (z_j')^{u_1+u_2+u_3}
\frac{(q/z'_j; q)_{\infty}(q^{-u_2} \Lambda_0/z'_j; q)_{\infty}}
{(q^{u_3+1}/z'_j; q)_{\infty} (\Lambda_0/z'_j; q)_{\infty}} \cr
& \times 
\prod_{1 \leq i<j\leq N_+} z_i^{2\beta}
\left( 1 - \frac{z_j}{z_i} \right)\frac{(q^{1-\beta} z_j/z_i; q)_{\infty}}
{(q^{\beta} z_j/z_i; q)_{\infty}}
\prod_{1 \leq i<j \leq N_-} (z'_j)^{2\beta}
\left( 1 - \frac{z'_i}{z'_j} \right)
\frac{(q^{1-\beta} z'_i/z'_j; q)_{\infty}}
{(q^{\beta} z'_i/z'_j; q)_{\infty}} \cr
& \times  \prod_{i=1}^{N_+} \prod_{j=1}^{N_-}
(z'_j)^{2\beta}
\left(1 - \frac{z_i}{z'_j} \right)
\frac{(q^{1-\beta} z_i/z'_j; q)_{\infty}}{(q^{\beta} z_i/z'_j; q)_{\infty}}.
\end{split}
\ee
Here the constant $\mathcal{V}_0(\Lambda_0)$ is given by
\be
\begin{split}
\mathcal{V}_0(\Lambda_0)&:=
\Lambda_0^{1/(2\beta) u_1 u_2 + u_2 N_+ }
q^{(u_1+u_2+2\beta N_+)u_3(u_3+1)/(2\beta)}  \cr
& \qq \times \exp\left( -\sum_{n=1}^{\infty} \frac{1}{n}
\frac{(q^{-nu_3}-1)(q^{-nu_2}-1)}{(1-q^n)(1-t^n)(1+p^n)}
\Lambda_0^n \right).
\end{split}
\ee
The ``range'' $\mathcal{C}_{N_+}([0,\Lambda_0])$
of the Jackson integral for $\{z_i\}_{i=1,2,\dotsc, N_+}$
are the interval $[0, \Lambda_0]$ with the additional condition
\be
0 \leq |z_{N_+}| < |z_{N_+-1} | < \dotsm < |z_2|<|z_1| \leq |\Lambda_0|.
\ee
Recall that we have assumed $|q|<1$. Hence, this condition means that, for 
a parameterization $z_i = \Lambda_0 q^{k_i}$ in the Jackson integral, the non-negative
integers $k_i$ are summed over with
\be
0 \leq k_1 < k_2 < \dotsm < k_{N_+}.
\ee
Similarly, the ``range'' $\mathcal{C}'_{N_-}([1,\infty])$
is chosen such that
\be
1 \leq |z'_1| < |z'_2| < \dotsm < |z'_{N_-}| < \infty.
\ee
It means that $z'_j = q^{-k'_j}$ with $0 \leq k'_1 < k'_2 < \dotsm <k'_{N_-}$.

By the following coordinate transformation:
\be
z_i = \Lambda_0 x_i, \ (i=1,2, \dotsc, N_+), \qq
z'_j = 1/y_j, \ (j=1,2,\dotsc, N_-),
\ee
we have
\bel{BL0}
\begin{split}
\mathcal{B}(\Lambda_0) &= \mathcal{V}_0'(\Lambda_0) \, (N_+)! (N_-)! 
\int_{C_{N_+}([0,1])} \de_q^{N_+} x \int_{C_{N_-}([0,1])} \de_q^{N_-} y \cr
& \times \prod_{i=1}^{N_+} x_i^{u_1}
\frac{(q x_i; q)_{\infty}}{(q^{u_2+1} x_i; q)_{\infty}}
\prod_{1 \leq i<j \leq N_+} x_i^{2\beta}
\left(1 - \frac{x_j}{x_i} \right) \frac{(q^{1-\beta} x_j/x_i; q)_{\infty}}
{(q^{\beta} x_j/x_i; q)_{\infty}} \cr
& \times \prod_{j=1}^{N_-} y_j^{u_4}
\frac{(q y_j; q)_{\infty}}{(q^{u_3+1} y_j; q )_{\infty}}
\prod_{1 \leq i<j\leq N_-} y_i^{2\beta}
\left( 1 - \frac{y_j}{y_i} \right) 
\frac{(q^{1-\beta} y_j/y_i; q)_{\infty}}
{(q^{\beta} y_j/y_i; q)_{\infty}} \cr
& \times \prod_{i=1}^{N_+}
\frac{(q^{-u_3} \Lambda_0 x_i; q)_{\infty}}
{(\Lambda_0 x_i; q)_{\infty}}
\prod_{j=1}^{N_-}
\frac{(q^{-u_2} \Lambda_0 y_j; q)_{\infty}}
{(\Lambda_0 y_j; q)_{\infty}}
\prod_{i=1}^{N_+} \prod_{j=1}^{N_-}
(1 - \Lambda_0 x_i y_j )
\frac{(q^{1-\beta} \Lambda_0 x_iy_j; q)_{\infty}}
{(q^{\beta} \Lambda_0 x_i y_j; q)_{\infty}},
\end{split}
\ee
where
\be
\begin{split}
\mathcal{V}_0'(\Lambda_0)
&= \mathcal{V}_0(\Lambda_0) \, \Lambda_0^{N_+(u_1+1) + \beta N_+(N_+-1)} \cr
&=\Lambda_0^{1/(2\beta) u_1 u_2 + N_+(u_1+u_2+1) + \beta N_+ (N_+-1)}
q^{(u_1+u_2+2\beta N_+)u_3(u_3+1)/(2\beta)}  \cr
& \qq \times \exp\left( -\sum_{n=1}^{\infty} \frac{1}{n}
\frac{(q^{-nu_3}-1)(q^{-nu_2}-1)}{(1-q^n)(1-t^n)(1+p^n)}
\Lambda_0^n \right).
\end{split}
\ee
The Jackson integrals in \eqref{BL0} are taken for the range $[0,1]$ under the following conditions
\be
0 \leq |x_{N_+}|< |x_{N_+-1}| < \dotsm < |x_2|<|x_1| \leq 1,
\ee
\be
0 \leq |y_{N_-}|< |y_{N_--1}| < \dotsm < |y_2|<|y_1| \leq 1.
\ee
Let
\bel{B0N}
\mathcal{B}_0(\Lambda_0):= \mathcal{B}(\Lambda_0)/\mathcal{V}_0'(\Lambda_0). 
\ee
At $\Lambda_0=0$, $\mathcal{B}_0(\Lambda_0)$
factorizes into a product of two $q$-deformed Selberg integrals:
\be
\mathcal{B}_0(0) = S^{(q)}_{N_+}(u_1, u_2, \beta) S^{(q)}_{N_-}(u_4, u_3, \beta),
\ee
where
\bel{SNq0}
\begin{split}
& S^{(q)}_{N}(u_1, u_2, \beta) \cr
&= N! \int_{C_N([0,1])}
\de^N_q z \prod_{i=1}^N z_i^{u_1}
\frac{(q z_i; q)_{\infty}}{(q^{u_2+1} z_i; q)_{\infty}}
\prod_{1 \leq i<j \leq N} z_i^{2\beta}
\left(1 - \frac{z_j}{z_i} \right)
\frac{(q^{1-\beta} z_j/z_i; q)_{\infty}}
{(q^{\beta} z_j/z_i; q)_{\infty}} \cr
&= N! (1-q)^N \sum_{0 \leq k_1 < k_2 < \dotsm < k_N }
\prod_{i=1}^N q^{(u_1+1) k_i}
\frac{(q^{k_i+1}; q)_{\infty}}{(q^{k_i+u_2+1}; q)_{\infty}}
\prod_{1 \leq i<j \leq N}
q^{2\beta k_i}
( 1 - q^{k_j-k_i})
\frac{(q^{k_j-k_i+1-\beta}; q)_{\infty}}
{(q^{k_j-k_i+\beta}; q)_{\infty}}.
\end{split}
\ee

When $\beta$ is a positive integer, \eqref{SNq0}
becomes
\bel{SNq0int}
\begin{split}
& S_N^{(q)}(u_1,u_2,\beta)  \cr
&= \int_{[0,1]^N} \de^N_q z \prod_{i=1}^N z_i^{u_1}
\frac{(q z_i; q)_{\infty}}{(q^{u_2+1} z_i; q)_{\infty}}
\prod_{1 \leq i<j \leq N} \left\{ (z_i - z_j) 
\prod_{\ell=0}^{2\beta-2} ( z_i - q^{\ell+1-\beta} z_j ) \right\}.
\end{split}
\ee
This integral \eqref{SNq0int} is evaluated as
\bel{SNq0f}
S_N^{(q)}(u_1,u_2,\beta) =N! \, q^{A_N(u_1,\beta)} \prod_{j=1}^N
\frac{\Gamma_q(u_1+1+(N-j) \beta) \Gamma_q(u_2+1+(N-j)\beta)
\Gamma_q(j\beta)}
{\Gamma_q(u_1+u_2+2+(2N-j-1) \beta) \Gamma_q(\beta)},
\ee
where $\Gamma_q(x)$ is the $q$-Gamma function and
\bel{AN}
A_N(u_1,\beta) = \frac{1}{2} N(N-1) (u_1+1) \beta 
+ \frac{1}{3} N(N-1)(N-2) \beta^2 ,\qq
(\mbox{$\beta$: a positive integer}).
\ee
We expect that \eqref{SNq0f} also holds when $\beta$
is not a positive integer with a modification of $A_{N}(u_1,\beta)$
from \eqref{AN}. 

For example, for positive $u_1$ and $u_2$,
the small $q$ behavior of $S_N^{(q)}$ is given by $N!\,  q^{A_N}+\dotsm$.
When $0 < \beta <2$, the leading contribution in the sum \eqref{SNq0}
comes from the term with 
$k_i=(i-1)$.
Hence, it seems that in this case, 
\be
A_N(u_1, \beta) = \frac{1}{2} N(N-1) (u_1+1) 
+ \frac{1}{3} N(N-1)(N-2) \beta, \qq ( 0 < \beta < 2).
\ee

%%%%%%%%%%%%%%%%%%%%%%%%%%%%%%%%%%%%%%%%%%%%%%%%%%%
\subsection{Remark: $q\rightarrow 1$ limit}

In this subsection, we comment on the $q \rightarrow 1$ limit of various objects in previous subsections.

For notational simplicity, let us introduce the following functions:
\bel{DNq}
D_N^{(q)}(u_1, u_2, \beta; z)
= \prod_{j=1}^N z_j^{u_1} 
\frac{(q z_j; q)_{\infty}}{(q^{u_2+1} z_j; q)_{\infty}}
\prod_{1 \leq i<j \leq N} z_i^{2\beta-1}
\frac{(q^{1-\beta} z_j/z_i; q)_{\infty}}{(q^{\beta} z_j/z_i; q)_{\infty}}
(z_i - z_j),
\ee
\be
\begin{split}
& F^{(q)}_{N_+, N_-}(u_2, u_3, \beta, \Lambda_0; x, y) \cr
&= \prod_{i=1}^{N_+}
\frac{(\Lambda_0 q^{-u_3} x_i; q)_{\infty}}
{(\Lambda_0 x_i; q)_{\infty}}
\prod_{j=1}^{N_-}
\frac{(\Lambda_0 q^{-u_2} y_j; q)_{\infty}}
{(\Lambda_0 y_j; q)_{\infty}}
\prod_{i=1}^{N_+} \prod_{j=1}^{N_-}
\frac{
(q^{1-\beta} \Lambda_0 x_i y_j; q)_{\infty}}
{
(q^{\beta} \Lambda_0 x_i y_j; q)_{\infty}}
(1- \Lambda_0 x_i y_j).
\end{split}
\ee

In the $q \rightarrow 1$ limit, these objects behave
as follows:
\be
\lim_{q \rightarrow 1}
D_N^{(q)}(u_1, u_2, \beta; z)
= \prod_{j=1}^{N} z_j^{u_1} (1-z_j)^{u_2}
\prod_{1 \leq i<j \leq N} (z_i -z_j)^{2\beta}.
\ee
\be
\begin{split}
& \lim_{q \rightarrow 1}
F^{(q)}_{N_+, N_-}(u_2, u_3, \beta, \Lambda_0; x, y) \cr
&= \prod_{i=1}^{N_+} ( 1 - \Lambda_0 x_i)^{u_3}
\prod_{j=1}^{N_-}( 1 - \Lambda_0 y_j)^{u_2}
\prod_{i=1}^{N_+} \prod_{j=1}^{N_-}
( 1 - \Lambda_0 x_i y_j)^{2\beta}.
\end{split}
\ee
Hence,  \eqref{BL0} goes to Eq.(2.8) of \cite{IO5}.

Using \eqref{DNq}, the $q$-deformed Selberg integral
\eqref{SNq0} can be written as
\be
S_N^{(q)}(u_1,u_2, \beta)
= N! \int_{C_N([0,1])} \de^N_q z\, D^{(q)}_N(u_1,u_2, \beta; z).
\ee
In the $q \rightarrow 1$ limit, the $q$-Selberg integral \eqref{SNq0} with
\eqref{SNq0f} goes to the ordinary Selberg integral:
\be
\begin{split}
\lim_{q \rightarrow 1} S_N^{(q)}(u_1, u_2, \beta)
&= \int_0^1 \de^N z\,
\prod_{j=1}^N z_j^{u_1} (1-z_j)^{u_2}
\prod_{1 \leq i<j \leq N}
| z_i - z_j |^{2\beta} \cr
&= \prod_{j=1}^N 
\frac{\Gamma(u_1+1+(N-j) \beta) \Gamma(u_2+1+(N-j)\beta)
\Gamma(1+j\beta)}
{\Gamma(u_1+u_2+2+(2N-j-1) \beta) \Gamma(1+\beta)}.
\end{split}
\ee
Here we have used $\lim_{q \rightarrow 1} \Gamma_q(x) = \Gamma(x)$ and
\be
N! \prod_{j=1}^N \frac{\Gamma(j \beta)}{\Gamma(\beta)}
= \prod_{j=1}^N \frac{\Gamma(1+j \beta)}{\Gamma(1+\beta)}.
\ee

The $q \rightarrow 1$ limit is related to the 2d CFT and the 4d gauge theory on the flat space.
While root of unity limits of $q$ are related to 2d supersymmetric/parafermionic theories and
the 4d gauge theories on ALE spaces \cite{BF1105,NT1106,BBB1106,BMT,
I1110,AT1110,BW1205,BM1210,BBT1211,ABT1306,IOY2,spo1409,yos1512}.

%%%%%%%%%%%%%%%%%%%%%%%%%%%%%%%%%%%%%%%%%%%%%
\section{5d Nekrasov partition function}

In this section, we briefly review the Nekrasov partition function
on $\mathbb{R}^{3,1} \times S^1$. We denote the radius of $S^1$ by $R$.
The five dimensional $SU(N)$ Nekrasov partition function with
$N_f=2N$ fundamental matters can be found in \cite{nek0206,NO0306} (see also \cite{NY0505}).
We follow the notation of \cite{AY1004} and consider the $N=2$ case.

The instanton part of the five dimensional $SU(2)$ Nekrasov partition function
with $N_f=4$ fundamental matters is given by
\bel{Z2inst}
Z_2^{\mathrm{inst}}(Q; \Lambda)
= \sum_{\lambda, \mu} Z^{(\pm)}_{\lambda, \mu} 
\left( \frac{\Lambda_{\alpha}^{\pm}}{v^2}
\right)^{|\lambda|+|\mu|},
\ee
where
\be
Z_{\lambda, \mu}^{(\pm)}
= \frac{\prod_{i=1}^2 N_{\lambda,(0)}(vQ_1/Q_i^{\pm})
N_{\mu, (0)}(vQ_2/Q_i^{\pm}) N_{(0),\lambda}(vQ_i^{\mp}/Q_1)
N_{(0),\mu}(v Q_i^{\mp}/Q_2)}
{N_{\lambda, \lambda}(1) N_{\mu, \mu}(1) N_{\lambda, \mu}(Q_1/Q_2)
N_{\mu, \lambda}(Q_2/Q_1)}.
\ee
Here $v=(q/t)^{1/2} = p^{1/2}$. 
The summation in \eqref{Z2inst} is over a pair of partitions $(\lambda, \mu)$.
The function $N_{\lambda \mu}(Q)$ is defined by
\be
\begin{split}
N_{\lambda \mu}(Q)
&:= \prod_{(i,j) \in \lambda} \bigl( 1 - Q q^{\lambda_i-j}
t^{\mu'_j -i+1} \bigr)
\prod_{(i,j) \in \mu}
\bigl( 1 - Q q^{-\mu_i+j-1} t^{-\lambda'_j+i} \bigr) \cr
&= \prod_{(i,j) \in \mu}
\bigl( 1 - Q q^{\lambda_i - j} t^{\mu'_j -i+1} \bigr)
\prod_{(i,j) \in \lambda}
\bigl( 1 - Q q^{-\mu_i+j-1} t^{-\lambda'_j+i} \bigr).
\end{split}
\ee
Here $\lambda'$ is the conjugate partition of $\lambda$.

The parameters in \eqref{Z2inst} are $q$, $t$, $\Lambda$, 
$Q_1$, $Q_2$, $Q_1^{\pm}$ and $Q_2^{\pm}$. $q$ and $t$
are related to the $\Omega$-background parameters $\epsilon_1$ and $\epsilon_2$:
\bel{qtepsilon}
q = \ex^{R \epsilon_2}, \qq t = \ex^{-R \epsilon_1}.
\ee
The parameters $Q_1$ and $Q_2$ are related to the vev of the 
adjoint scalar in the 4d theory:
\bel{Q12a}
v Q_1 = \ex^{Ra}, \qq v Q_2 = \ex^{-Ra},
\ee
while $Q_i^{\pm}$ are related to the mass of fundamental matters:
\bel{Qpmmass}
Q_1^+ = \ex^{-R m_1}, \qq
Q_2^+ = \ex^{-R m_2}, \qq
Q_1^- = \ex^{-R m_3}, \qq
Q_2^- = \ex^{-R m_4}.
\ee
The expansion parameters $\Lambda_{\alpha}^{\pm}$ are defined by
\be
\Lambda_{\alpha}^{\pm} = \Lambda^4 \left(
\frac{Q_1^{\pm} Q_2^{\pm}}{Q_1^{\mp} Q_2^{\mp}}\right)^{1/2} = \Lambda^4\, 
\ex^{\pm (1/2)R(m_3+m_4-m_1-m_2)}.
\ee
Hence the relation between $Z^{(\pm)}_{\lambda, \mu}$ is given by
\be
Z^{(+)}_{\lambda, \mu} = \ex^{(|\lambda|+|\mu|) R(m_1+m_2-m_3-m_4)}
Z^{(-)}_{\lambda, \mu}.
\ee

%%%%%%%%%%%%%%%%%%%%%%%%%%%%%%%%%%%%%%%%%%
\subsection{First order: $Z^{(\pm)}_{(1),(0)}$ and $Z^{(\pm)}_{(0),(1)}$}

Using \eqref{qtepsilon}, \eqref{Q12a} and \eqref{Qpmmass}, 
$Z^{\pm}_{\lambda , \mu}$ with $|\lambda|+|\mu|=1$ can be written as
\bel{Zp10}
Z^{(+)}_{(1),(0)}
= \frac{(1-\ex^{R(a+m_1)})(1-\ex^{R(a+m_2)})
(1-\ex^{-R(a+m_3)})(1-\ex^{-R(a+m_4)})}
{(1-\ex^{-R\epsilon_1})(1-\ex^{-R\epsilon_2})
(1-\ex^{2Ra})(1-\ex^{-R(2a+\epsilon)})},
\ee
\bel{Zp01}
Z^{(+)}_{(0),(1)}
= \frac{(1-\ex^{-R(a-m_1)})(1-\ex^{-R(a-m_2)})
(1-\ex^{R(a-m_3)})(1-\ex^{R(a-m_4)})}
{(1-\ex^{-R\epsilon_1})(1-\ex^{-R\epsilon_2})
(1-\ex^{-2Ra})(1-\ex^{R(2a-\epsilon)})}.
\ee
\bel{Zm10}
Z^{(-)}_{(1),(0)}
= \frac{(1-\ex^{-R(a+m_1)})(1-\ex^{-R(a+m_2)})
(1-\ex^{R(a+m_3)})(1-\ex^{R(a+m_4)})}
{(1-\ex^{-R\epsilon_1})(1-\ex^{-R\epsilon_2})
(1-\ex^{2Ra})(1-\ex^{-R(2a+\epsilon)})},
\ee
\bel{Zm01}
Z^{(-)}_{(0),(1)}
= \frac{(1-\ex^{R(a-m_1)})(1-\ex^{R(a-m_2)})
(1-\ex^{-R(a-m_3)})(1-\ex^{-R(a-m_4)})}
{(1-\ex^{-R\epsilon_1})(1-\ex^{-R\epsilon_2})
(1-\ex^{-2Ra})(1-\ex^{R(2a-\epsilon)})}.
\ee
Here $\epsilon = \epsilon_1+\epsilon_2$.

In the $R \rightarrow 0$ limit, these terms reproduce the 4d results:
\be
\lim_{R \rightarrow 0}
Z^{(\pm)}_{(1),(0)}
= \frac{(a+m_1)(a+m_2)(a+m_3)(a+m_4)}{2a(2a+\epsilon)g_s^2},
\ee
\be
\lim_{R \rightarrow 0}
Z^{(\pm)}_{(0),(1)}
= \frac{(a-m_1)(a-m_2)(a-m_3)(a-m_4)}{2a(2a-\epsilon)g_s^2}.
\ee

%%%%%%%%%%%%%%%%%%%%%%%%%%%%%%%%%%%
\section{$\Lambda_0$ expansion of $q$-block}

In this section, we study the expansion of the modified blocks in the $\Lambda_0$ parameter.

%%%%%%%%%%%%%%%%%%%%%%%%%%
\subsection{Method of calculation in $q$-block}

In order to compare the $q$-block \eqref{B0N} with the 5d Nekrasov function \eqref{Z2inst}, 
let us introduce the following function
\bel{AL0}
\mathcal{A}(\Lambda_0)
:= \frac{\mathcal{B}_0(\Lambda_0)}{\mathcal{B}_0(0)}
= \left\langle\!\!\left\langle
F^{(q)}_{N_+, N_-}(u_2,u_3, \beta, \Lambda_0; x, y)
\right\rangle\!\!\right\rangle 
=  1 + \sum_{n=1}^{\infty} \Lambda_0^n \mathcal{A}_n.
\ee
Here
\be
\bigl\langle \!\bigl\langle f(x,y ) 
\bigr\rangle \! \bigr\rangle=\bigl\langle \bigl\langle f(x,y) \bigr\rangle_+
\bigr\rangle_-,
\ee
and $\langle \ \rangle_{\pm}$ is 
the average with respect to the $q$-deformed Selberg
measure:
\be
\langle f(x) \rangle_+
= \frac{1}{S_{N_+}^{(q)}(u_1, u_2, \beta)}
\int_{C_{N_+}([0, 1])} \de^{N_+}_q x\, D_{N_+}^{(q)}
(u_1, u_2, \beta; x)\, f(x),
\ee
\be
\langle f(y) \rangle_-
= \frac{1}{S_{N_-}^{(q)}(u_4, u_3, \beta)}
\int_{C_{N_-}([0, 1])} \de^{N_-}_q y\, D_{N_-}^{(q)}
(u_4, u_3, \beta; y)\, f(y).
\ee

Note that
\be
\begin{split}
F_{N_+, N_-}^{(q)} 
&= \exp\left[ - \sum_{k=1}^{\infty}
\frac{\Lambda_0^k}{k} 
\frac{(1-t^k)}{(1-q^k)}
\left\{ \left( p_k(x) - \frac{1-q^{-u_2k} }{1-t^k} \right)
p_k(y) \right. \right. \cr 
& \qq \qq \qq \qq \qq \left. \left.+ p_k(x) \left( (q/t)^{k}\, p_k(y)
- \frac{1-q^{-u_3k}}{1-t^k} \right) \right\} \right],
\end{split}
\ee
where $p_k$ denotes the power sum:
\be
p_k(x) = \sum_{i=1}^{N_+} x_i^k, \qq p_k(y)
= \sum_{j=1}^{N_-} y_j^k.
\ee

We conjecture that the Kaneko's formula \eqref{MacAV} also holds for
the contour $C_{N_{\pm}}([0,1])$. We have checked it for small values of $N$
and $\beta$.

Then, we can calculate $\mathcal{A}_n$ by using the average of the Macdonald polynomial
$P_{\lambda}$:
\be
\begin{split}
\langle P_{\lambda}(x; q,t) \rangle_+
&= \frac{(t^{N_+})_{\lambda}^{(q,t)}(q^{u_1+1} t^{(N_+-1)})_{\lambda}^{(q,t)}}
{h_{\lambda}(q,t) (q^{u_1+u_2+2} t^{2(N_+-1)})_{\lambda}^{(q,t)}} \cr
&= \prod_{(i,j) \in \lambda}
\frac{(t^{i-1} -q^{j-1} t^{N_+} )(t^{i-1} - q^{u_1+j} t^{N_+-1})}
{(1-q^{\lambda_i - j } t^{\lambda'_j-i+1})
(t^{i-1} - q^{u_1+u_2+j+1} t^{2N_+-2})},
\end{split}
\ee
\be
\begin{split}
\langle P_{\lambda}(y; q,t) \rangle_-
&= \frac{(t^{N_-})_{\lambda}^{(q,t)}(q^{u_4+1} t^{(N_--1)})_{\lambda}^{(q,t)}}
{h_{\lambda}(q,t) (q^{u_3+u_4+2} t^{2(N_--1)})_{\lambda}^{(q,t)}} \cr
&= \prod_{(i,j) \in \lambda}
\frac{(t^{i-1} -q^{j-1} t^{N_-} )(t^{i-1} - q^{u_4+j} t^{N_--1})}
{(1-q^{\lambda_i - j } t^{\lambda'_j-i+1})
(t^{i-1} - q^{u_3+u_4+j+1} t^{2N_--2})}.
\end{split}
\ee

%%%%%%%%%%%%%%%%%%%%%%%%%%%%%%%%%%%%%%%%%%%%%%%%%%%%
\subsection{Parameter dictionary}

In \cite{IO5}, we have used the following identification of $\beta$ with
the $\Omega$-background parameters $\epsilon_1$ and $\epsilon_2$:
\be
\epsilon_1 = \sqrt{\beta} g_s, \qq 
\epsilon_2 = - \frac{1}{\sqrt{\beta}} g_s, \qq
\epsilon = \epsilon_1 + \epsilon_2 = \left( \sqrt{\beta} - \frac{1}{\sqrt{\beta}}
\right) g_s.
\ee
In this choice, it holds that  $\epsilon_1 \epsilon_2 = - g_s^2$.

With this identification, \eqref{qtepsilon} yields the following relations:
\be
q= \ex^{R \epsilon_2} = \ex^{-R g_s/\sqrt{\beta}}, \qq
t = q^{\beta} = \ex^{-R \epsilon_1} = \ex^{-R g_s \sqrt{\beta}}, \qq
v= (q/t)^{1/2} = \ex^{(1/2)R \epsilon}.
\ee

The momentum conservation condition \eqref{MCC} can be rewritten as
\be
q^{u_1+u_2+u_3+u_4+2} \, t^{2(N_++N_-)-2} = 1.
\ee

The 2d/4d dictionary used in \cite{IO5} now converted to the
following 2d/5d dictionary:
\be
t^{N_+} = \ex^{-R(a-m_2)}, \qq
t^{N_-} = \ex^{R(a+m_3)}, 
\ee
\be
q^{u_1} = \ex^{-R(m_2-m_1+\epsilon)}, \qq
q^{u_2} = \ex^{-R(m_1+m_2)},
\ee
\be
q^{u_3} = \ex^{-R(m_3+m_4)}, \qq
q^{u_4} = \ex^{-R(m_3-m_4+ \epsilon)}.
\ee

Some useful relations are given by
\be
q^{u_1+1}t^{-1} = \ex^{-R(m_2-m_1)}, \qq
q^{u_2} = \ex^{-R(m_1+m_2)}, 
\ee
\be
q^{u_3} = \ex^{-R(m_3+m_4)}, \qq
q^{u_4+1} t^{-1} = \ex^{-R(m_3-m_4)},
\ee
\be
q^{u_1+u_2+1} t^{2N_+-1} = \ex^{-2Ra}, \qq
q^{u_3+u_4+1} t^{2N_--1} = \ex^{2Ra}.
\ee

%%%%%%%%%%%%%%%%%%%%%%%%%%%%%%%%%%%%%%%%%%%%%%%%%%%%%%%%%%%%%
\subsection{First order: $\mathcal{A}_1$}

The first order term in the $\Lambda_0$-expansion \eqref{AL0} is given by
\be
\begin{split}
\mathcal{A}_1 
&= - \frac{(1-t)}{(1-q)}(1+v^2) \langle p_1(x) \rangle_+ 
\langle p_1(y) \rangle_- \cr
& + \frac{1-q^{-u_3}}{1-q} \langle p_1(x) \rangle_+
+ \frac{1-q^{-u_2}}{1-q} \langle p_1(y) \rangle_-.
\end{split}
\ee
Note that the following identity holds for any $A$ and $v$:
\be
\frac{1-v^2A}{1-A} + \frac{1-v^2 A^{-1}}{1-A^{-1}} = 1+v^2
\ee
Specializing this relation by setting $A$ to
\be
A = q^{u_1+u_2+1} \, t^{2N_+-1}, \qq A^{-1} = q^{u_3+u_4+1} \, t^{2N_--1},
\ee
we have
\bel{1v2}
1 + v^2 = \frac{1-q^{u_1+u_2+2}\,  t^{2N_+-2}}{1-q^{u_1+u_2+1} \, t^{2N_+-1}}
+ \frac{1 - q^{u_3+u_4+2}\,  t^{2N_--2}}{1-q^{u_3+u_4+1}\,  t^{2N_--1}}.
\ee

With help of \eqref{1v2}, a non-trivial decomposition of $\mathcal{A}_1$ is obtained:
\be
\mathcal{A}_1 = \mathcal{A}_{(1),(0)} + \mathcal{A}_{(0),(1)},
\ee
where
\bel{A10mid}
\mathcal{A}_{(1),(0)}
= \left\{ - \frac{(1-t)(1-q^{u_1+u_2+2} \, t^{2N_+-2})}
{(1-q)(1-q^{u_1+u_2+1}\, t^{2N_+-1})} \langle p_1(x) \rangle_+
+ \frac{(1-q^{-u_2})}{(1-q)} \right\} \langle p_1(y) \rangle_-,
\ee
\bel{A01mid}
\mathcal{A}_{(0),(1)}
= \langle p_1(x) \rangle_+
\left\{ - \frac{(1-t)(1-q^{u_3+u_4+2}\, t^{2N_--2})}
{(1-q)(1-q^{u_3+u_4+1}\, t^{2N_--1})} \langle p_1(y) \rangle_-
+ \frac{(1-q^{-u_3})}{(1-q)} \right\}.
\ee

The averages for the Macdonald polynomial with $\lambda =(1)$ are given by
\be
\langle P_{(1)}(x; q,t) \rangle_+
=\langle p_1(x) \rangle_+
= \frac{(1-t^{N_+})(1-q^{u_1+1} \, t^{N_+-1})}{(1-t)(1-q^{u_1+u_2+2}\, 
t^{2N_+-2})},
\ee
\be
\langle P_{(1)}(y; q,t) \rangle_-
=\langle p_1(y) \rangle_-
= \frac{(1-t^{N_-})(1-q^{u_4+1} \, t^{N_--1})}{(1-t)(1-q^{u_3+u_4+2}\, 
t^{2N_--2})}.
\ee
By substituting these expressions into factors in curly bracket of \eqref{A10mid} and 
\eqref{A01mid}, we can see that
\be
\begin{split}
& \left\{ - \frac{(1-t)(1-q^{u_1+u_2+2} \, t^{2N_+-2})}
{(1-q)(1-q^{u_1+u_2+1}\, t^{2N_+-1})} \langle p_1(x) \rangle_+
+ \frac{(1-q^{-u_2})}{(1-q)} \right\} \cr
&= - \frac{(1-t^{N_+})(1-q^{u_1+1} \, t^{N_+-1})}{(1-q)(1-q^{u_1+u_2+1}\, t^{2N_+-1})}
+ \frac{(1-q^{-u_2})}{(1-q)} \cr
&=-q^{u_1+u_2+1} \, t^{2N_+-1}
\frac{(1-q^{-u_2} \, t^{-N_+})(1-q^{-u_1-u_2-1} \, t^{1-N_+})}
{(1-q)(1-q^{u_1+u_2+1} \, t^{2N_+-1})},
\end{split}
\ee
\be
\begin{split}
& \left\{ - \frac{(1-t)(1-q^{u_3+u_4+2} \, t^{2N_--2})}
{(1-q)(1-q^{u_3+u_4+1}\, t^{2N_--1})} \langle p_1(y) \rangle_-
+ \frac{(1-q^{-u_3})}{(1-q)} \right\} \cr
&= - \frac{(1-t^{N_-})(1-q^{u_4+1} \, t^{N_--1})}{(1-q)(1-q^{u_3+u_4+1}\, t^{2N_--1})}
+ \frac{(1-q^{-u_3})}{(1-q)} \cr
&=-q^{u_3+u_4+1} \, t^{2N_--1}
\frac{(1-q^{-u_3} \, t^{-N_-})(1-q^{-u_3-u_4-1} \, t^{1-N_-})}
{(1-q)(1-q^{u_3+u_4+1} \, t^{2N_--1})}.
\end{split}
\ee
Consequently, we have explicit form of $\mathcal{A}_{(1),(0)}$ and $\mathcal{A}_{(0),(1)}$:
\bel{A10q}
\begin{split}
& \mathcal{A}_{(1),(0)} \cr
&=-q^{u_1+u_2+1} \, t^{2N_+-1}
\frac{(1-q^{-u_2} \, t^{-N_+})(1-q^{-u_1-u_2-1} \, t^{1-N_+})
(1-t^{N_-})(1-q^{u_4+1} \, t^{N_--1})}
{(1-q)(1-t)(1-q^{u_1+u_2+1} \, t^{2N_+-1})(1-q^{u_3+u_4+2} \, t^{2N_--2})},
\end{split}
\ee
\bel{A01q}
\begin{split}
& \mathcal{A}_{(0),(1)} \cr
&=-q^{u_3+u_4+1} \, t^{2N_--1}
\frac{(1-q^{-u_3} \, t^{-N_-})(1-q^{-u_3-u_4-1} \, t^{1-N_-})(1-t^{N_+})
(1-q^{u_1+1}\, t^{N_+-1})}
{(1-q)(1-t)(1-q^{u_3+u_4+1} \, t^{2N_--1})(1-q^{u_1+u_2+2} \, t^{2N_+-2})}.
\end{split}
\ee

%%%%%%%%%%%%%%%%%%%%%%%%%%%%%%%%%%%%%%%%%
\subsubsection{In terms of parameters of gauge theory}

Let us rewrite the parameters in \eqref{A10q} and \eqref{A01q} by
the gauge theory parameters.

Using
\be
q^{-u_2} = \ex^{R(m_1+m_2)}, \qq
q^{-u_3} = \ex^{R(m_3+m_4)}.
\ee
\be
t^{N_+} = \ex^{-R(a-m_2)}, \qq
q^{u_1+1} \, t^{N_+-1} = \ex^{-R(a-m_1)},
\ee
\be
q^{u_1+u_2+2} \, t^{2N_+-2} = \ex^{-R(2a-\epsilon)},
\ee
\be
t^{N_-} = \ex^{R(a+m_3)}, \qq
q^{u_4+1} t^{N_--1} = \ex^{R(a+m_4)},
\ee
etc., we have
\be
\begin{split}
\langle p_1(x) \rangle_+
&= \frac{(1-\ex^{-R(a-m_1)})(1-\ex^{-R(a-m_2)})}
{(1-\ex^{-R\epsilon_1})(1-\ex^{-R(2a-\epsilon)})},
\cr
\langle p_1(y) \rangle_-
&= \frac{(1-\ex^{R(a+m_3)})(1-\ex^{R(a+m_4)})}
{(1-\ex^{-R\epsilon_1})(1-\ex^{R(2a+\epsilon)})},
\end{split}
\ee
\bel{L1g}
- \frac{(1-t)(1-q^{u_1+u_2+2} \, t^{2N_+-2})}
{(1-q)(1-q^{u_1+u_2+1}\, t^{2N_+-1})} \langle p_1(x) \rangle_+
+ \frac{(1-q^{-u_2})}{(1-q)}
= \frac{(1-\ex^{R(a+m_1)})(1-\ex^{R(a+m_2)})}
{(1-\ex^{R\epsilon_2})(1-\ex^{2Ra})},
\ee
\bel{R1g}
- \frac{(1-t)(1-q^{u_3+u_4+2} \, t^{2N_--2})}
{(1-q)(1-q^{u_3+u_4+1} \, t^{2N_--1})} \langle p_1(y) \rangle_-
+ \frac{(1-q^{-u_3})}{(1-q)} 
= \frac{(1-\ex^{-R(a-m_3)})(1-\ex^{-R(a-m_4)})}
{(1-\ex^{R\epsilon_2})(1-\ex^{-2Ra})}.
\ee
Thus, we finally have the following expressions:
\bel{A10}
\mathcal{A}_{(1),(0)}
= \frac{(1-\ex^{R(a+m_1)})(1-\ex^{R(a+m_2)})(1-\ex^{R(a+m_3)})(1-\ex^{R(a+m_4)})}
{(1-\ex^{R\epsilon_2})(1-\ex^{-R\epsilon_1})
(1-\ex^{2Ra})(1-\ex^{R(2a+\epsilon)})},
\ee
\bel{A01}
\mathcal{A}_{(0),(1)}
= \frac{(1-\ex^{-R(a-m_1)})(1-\ex^{-R(a-m_2)})(1-\ex^{-R(a-m_3)})(1-\ex^{-R(a-m_4)})}
{(1-\ex^{R\epsilon_2})(1-\ex^{-R\epsilon_1})(1-\ex^{-2Ra})(1-\ex^{-R(2a-\epsilon)})}.
\ee

\vspace{3mm}

\noindent
Remark: the following identity plays the crucial role in \eqref{L1g}:
\be
\begin{split}
& \ex^{-2Ra}(1-\ex^{R(a-m_1)})(1-\ex^{R(a-m_2)})
+ (1-\ex^{-2Ra})(1-\ex^{-R(m_1+m_2)}) \cr
&= (1 - \ex^{-R(a+m_1)})(1-\ex^{-R(a+m_2)}),
\end{split}
\ee
and similarly the following identity is used in \eqref{R1g}:
\be
\begin{split}
& \ex^{-2Ra}(1-\ex^{R(a+m_3)})(1-\ex^{R(a+m_4)})
+ (1-\ex^{-2Ra})(1-\ex^{R(m_3+m_4)}) \cr
&=(1-\ex^{-R(a-m_3)})(1-\ex^{-R(a-m_4)}).
\end{split}
\ee

%%%%%%%%%%%%%%%%%%%%%%%%%%%%%%%%%%%%%%%%%
\section{Comparison with 5d Nekrasov partition function}

In this section, we compare the modified $q$-block with the Nekrasov function.

We assume that the 2d/5d correspondence holds, i.e.,  
\bel{AtoZ}
\mathcal{A}(\Lambda_0)
= Z^{\mathrm{inst}}_2(Q; \Lambda).
\ee

By comparing \eqref{A10} with \eqref{Zp10} or \eqref{Zm10}, and \eqref{A01} with
\eqref{Zp01} or \eqref{Zm01}, 
we can check that
\bel{AZ10}
\mathcal{A}_{(1),(0)} = \left(\frac{t}{q^2} \right) \ex^{R(m_3+m_4)} Z^{(+)}_{(1),(0)}
= \left(\frac{t}{q^2}\right) \ex^{R(m_1+m_2)} Z^{(-)}_{(1),(0)},
\ee
\bel{AZ01}
\mathcal{A}_{(0),(1)} 
= \left(\frac{t}{q^2} \right)\ex^{R(m_3+m_4)} Z^{(+)}_{(0),(1)}
= \left(\frac{t}{q^2} \right)\ex^{R(m_1+m_2)} Z^{(-)}_{(0),(1)}.
\ee
Then we must have
\be
\Lambda_0 \mathcal{A}_{(1),(0)}
= \frac{\Lambda_{\alpha}^+}{v^2} Z^{(+)}_{(1),(0)}
= \frac{\Lambda_{\alpha}^-}{v^2} Z^{(-)}_{(1),(0)},
\ee
\be
\Lambda_0 \mathcal{A}_{(0),(1)}
= \frac{\Lambda_{\alpha}^+}{v^2} Z^{(+)}_{(0),(1)}
= \frac{\Lambda_{\alpha}^-}{v^2} Z^{(-)}_{(0),(1)}.
\ee

The relations \eqref{AZ10} and \eqref{AZ01} lead to 
the following identification of the expansion parameters:
\be
\Lambda_0
= q \, \ex^{-R(m_3+m_4)}\,  \Lambda_{\alpha}^+ 
= q \, \ex^{-R(m_1+m_2)}\,  \Lambda_{\alpha}^-
= q \, \ex^{-(1/2)R(m_1+m_2+m_3+m_4)} \, \Lambda^4.
\ee
This connection is also stated as follows:
\be
\Lambda_0 = q^{(1/2)(u_2+u_3)} \, \Lambda^4.
\ee

Using this identification, \eqref{AtoZ} decomposes into the
following identities:
\be
\mathcal{A}_k=
\left( \frac{t}{q^2} \right)^k \ex^{kR(m_3+m_4)}
\sum_{|\lambda|+|\mu|=k} Z^{(+)}_{\lambda, \mu}
= \left( \frac{t}{q^2} \right)^k \ex^{kR(m_1+m_2)}
\sum_{|\lambda|+|\mu|=k} Z^{(-)}_{\lambda, \mu}.
\ee
We have checked these identities up to  $k=4$. 
These are quite non-trivial relations even for the cases of low order $k$.
We expect that these hold for all $k$. Therefore,
this gives strong evidence of the 2d/5d correspondence \eqref{AtoZ}.

%%%%%%%%%%%%%%%%  acknowledgements %%%%%%%%%%%%%%%%%%%%%%
\section*{Acknowledgments}
We would like to thank Mikhail Bershtein and Yusuke Ohkubo for valuable discussions. 
This work was supported by
JSPS KAKENHI Grant Number 15K05059.
Support from JSPS/RFBR bilateral collaborations 
``Faces of matrix models in quantum field theory
and statistical mechanics'' (H.~I. and R.~Y.)  and
``Exploration of Quantum Geometry
via Symmetry and Duality'' (T.~O.) is gratefully appreciated.

%%%%%%%%%%%%%%%%%%%%%%%%%%%%%%%%%%%%%%%%%%%%%%%%%%%%%%%%%%%

%%%%%%%%%%%%%%%%%%%%%%%%%%%%%%%%%%%%%%%%%%%%%%%%%%%%%
\appendix

%%%%%%%%%%%%%%%%%%%%%%%%%%%%%%%%%%%%%%%%%%%%%
\section{Two-point correlation functions}
\label{TPCF}

In this section, we collect two-point functions utilized in Subsection \ref{qDCGR}.

The (radial ordered) two-point correlation functions for the $q$-deformed vertex operators \eqref{qVop} 
and the screening current \eqref{qSCp} are given by
\be
\begin{split}
& \langle V_{u_1}(z_1) V_{u_2}(z_2) \rangle \cr
&=\begin{cases}
\dis z_1^{u_1 u_2/(2\beta)} \exp\left( - \sum_{n=1}^{\infty}
\frac{1}{n} \frac{(q^{nu_1}-1)(q^{-nu_2}-1)}{(1-q^{-n})
(1-t^n)(1+p^n)} \left( \frac{z_2}{z_1} \right)^n \right), 
& |z_1|>|z_2|, \cr
 & \cr
\dis z_2^{u_1 u_2/(2\beta)} \exp\left( - \sum_{n=1}^{\infty}
\frac{1}{n} \frac{(q^{nu_2}-1)(q^{-nu_1}-1)}{(1-q^{-n})
(1-t^n)(1+p^n)} \left( \frac{z_1}{z_2} \right)^n \right), 
& |z_2|>|z_1|,
\end{cases}
\end{split}
\ee
\be
\begin{split}
\langle V_u(z_1) S_+(z_2) \rangle
&= \begin{cases}
\dis z_1^u \frac{(q z_2/z_1; q)_{\infty}}{(q^{u+1} z_2/z_1; q)_{\infty}},
& |z_1|>|z_2|, \cr
 &  \cr
\dis z_2^u \frac{(q^{-u} z_1/z_2; q)_{\infty}}{(z_1/z_2; q)_{\infty}},
& |z_2|>|z_1|,
\end{cases}
\end{split}
\ee
\be
\begin{split}
\langle S_+(z_1) S_+(z_2) \rangle 
&= \begin{cases}
\dis z_1^{2\beta} \left(1 - \frac{z_2}{z_1} \right)
\frac{(q^{1-\beta} z_2/z_1; q)_{\infty}}
{(q^{\beta} z_2/z_1; q)_{\infty}}, &
|z_1|>|z_2|, \cr
 & \cr
\dis z_2^{2\beta} \left(1 - \frac{z_1}{z_2} \right)
\frac{(q^{1-\beta} z_1/z_2; q)_{\infty}}
{(q^{\beta} z_1/z_2; q)_{\infty}}, &
|z_2|>|z_1|.
\end{cases}
\end{split}
\ee
Here
\be
(x; q)_{\infty} = \prod_{n=0}^{\infty} (1 - x q^n)
\ee
is the $q$-Pochhammer symbol.

%%%%%%%%%%%%%%%%%%%%%%%%%%%%%%%%%%%%%%%%%%%%%
\section{$q$-Selberg integral and Kaneko's formula}

In this section, we shortly summarize the $q$-Selberg integral and the Kaneko's formula \cite{kan96}.
 
%%%%%%%%%%%%%%%%%%%%%%%%%%%%%%%%%%%%%%%%%%%%%%%%%
\subsection{$q$-Selberg integral}

Using \eqref{DNq}, let us consider the following $q$-deformation
of the Selberg integral:
\bel{SNq}
S_{N}^{(q)}(u_1, u_2, \beta; \xi)
:= \int_{[0, \xi \infty]} \de_q^{N} z\,
D_N^{(q)}(u_1,u_2, \beta; z),
\ee
with
\be
\xi=(\xi_1, \xi_2, \dotsc, \xi_N) \in ( \mathbb{C}^*)^N.
\ee
Here
\be
\begin{split}
& \int_{[0, \xi \infty]} \de_q^N z\, f(z_1, z_2, \dotsc, z_N) \cr
&= (1-q)^N \left(\prod_{j=1}^N \xi_j\right)
\sum_{(s_1,s_2, \dotsc, s_N) \in \mathbb{Z}^N}
q^{s_1 + s_2 + \dotsm + s_N} f(\xi_1 q^{s_1}, \xi_2 q^{s_2},
\dotsc, \xi_N q^{s_N} ).
\end{split}
\ee

For certain value of the parameters $u_1$, $u_2$, $\beta$, $\xi$
such that the Jackson integral \eqref{SNq} converges,
the Aomoto's formula \cite{aom98} implies that (see also \cite{kan96})
\bel{aom}
\begin{split}
& S_N^{(q)}(u_1,u_2, \beta; \xi) \cr
&= q^{(1/2)N(N-1)^2 \beta}
\prod_{j=1}^N \xi_j^{u_1+2\beta(N-j) - N+2}
\frac{\vartheta(\xi_j q^{u_1+u_2+2+(\beta-1)(N-1)})
\vartheta(q^{u_2+1+(j-1)\beta}) \vartheta(q^{j \beta})}
{\vartheta(q^{u_1+u_2+(2\beta-1)(N-1)-(N-j)\beta})
\vartheta(\xi_j q^{u_2}) \vartheta(q^{\beta})} \cr
& \times \prod_{1 \leq i<j\leq N}
\frac{\vartheta(\xi_j/\xi_i)}{\vartheta(q^{\beta} \xi_j/\xi_i)}
\prod_{j=1}^N \frac{\Gamma_q(u_1+1+(j-1)\beta)
\Gamma_q(u_2+1+(j-1)\beta) \Gamma_q(j\beta)}
{\Gamma_q(u_1+u_2+2+\beta(N+j-2) )}.
\end{split}
\ee
Here $\vartheta(x)$ is the Jacobi elliptic function
\be
\vartheta(x) = (x;q)_{\infty} (q/x;q)_{\infty} (q;q)_{\infty},
\ee
and $\Gamma_q(x)$ is the $q$-Gamma function
\be
\Gamma_q(x) = (1-q)^{1-x} \frac{(q;q)_{\infty}}{(q^x; q)_{\infty}}.
\ee

%%%%%%%%%%%%%%%%%%%%%%%%%%%%%%%%%%%%%%%%%%%%%%
\subsection{Kaneko's formula}

Let $P_{\lambda}(z; q,t)$ be the Macdonald polynomial
for the variables $z=(z_1, z_2, \dotsc, z_N)$.
The average over the Macdonald polynomials is defined by
\be
\langle P_{\lambda}(z; q,t) \rangle
:= \frac{1}{S_N^{(q)}(u_1,u_2, \beta; \xi)}
\int_{[0, \xi \infty]} \de^N_q z\, P_{\lambda} (z; q,t)\, 
D_N^{(q)}(u_1, u_2, \beta; z).
\ee
It is given by \cite{kan96}
\bel{MacAV}
\begin{split}
\langle P_{\lambda}(z; q,t) \rangle
&= \frac{(t^{N})_{\lambda}^{(q,t)}(q^{u_1+1} t^{(N-1)})_{\lambda}^{(q,t)}}
{h_{\lambda}(q,t) (q^{u_1+u_2+2} t^{2(N-1)})_{\lambda}^{(q,t)}} \cr
&= \prod_{(i,j) \in \lambda}
\frac{(t^{i-1} -q^{j-1} t^{N_+} )(t^{i-1} - q^{u_1+j} t^{N-1})}
{(1-q^{\lambda_i - j } t^{\lambda'_j-i+1})
(t^{i-1} - q^{u_1+u_2+j+1} t^{2N-2})}.
\end{split}
\ee
Here
\be
(A)_{\lambda}^{(q,t)} = \prod_{s \in \lambda} \left( t^{\ell'(s)} - q^{a'(s)} A \right)
= \prod_{(i,j) \in \lambda} \left( t^{i-1} - q^{j-1} A \right),
\ee
\be
h_{\lambda}(q,t) = \prod_{s \in \lambda}\left( 1 - q^{a(s)} t^{\ell(s)+1} \right)
= \prod_{(i,j)\in \lambda} \left( 1 - q^{\lambda_i-j} t^{\lambda'_j-i+1} \right).
\ee
For a square $s=(i,j)$ in a partition $\lambda$, the arm-length, the leg-length, the arm-colength
and the leg-colength are respectively denoted by $a(s)$, $\ell(s)$, $a'(s)$ and $\ell'(s)$.

Note that the average \eqref{MacAV} does not depend on the choice
of $\xi$.

%%%%%%%%%%%%%%%%%%%%%%
\subsection{Special case}

When $\beta$ is a positive integer, $\beta=k$,  by choosing
$\xi=(1, q^{\beta}, q^{2\beta}, \dotsm, q^{(N-1)\beta})$, 
the Aomoto's formula \eqref{aom} reduces to
the Askey-Habsieger-Kadell's formula \cite{ask,hab,kad88} :
\bel{AHK}
\begin{split}
& \int_{[0,1]^N} \de_q z_1 \wedge \de_q z_2 \wedge \dotsm
\wedge \de_q z_N \, \prod_{i=1}^N z_i^{u_1} 
\frac{(q z_i; q)_{\infty}}{(q^{u_2+1 } z_i; q)_{\infty}}
\prod_{1 \leq i < j \leq N}
z_i^{2k} \frac{(q^{1-k} z_j/z_i; q)_{\infty}}
{(q^{1+k} z_j/z_i; q)_{\infty}} \cr
&= q^{A_N} \prod_{i=1}^N
\frac{\Gamma_q(u_1+1 + (N-i) k) \Gamma_q(u_2+1+(N-i)k)
\Gamma_q(1+ik)}
{\Gamma_q(u_1+u_2+2+(2N-i-1)k ) \Gamma_q(1+k)},
\end{split}
\ee
where
\be
A_N = \frac{1}{2} (u_1+1) k N(N-1)
+ \frac{1}{3} k^2 N(N-1)(N-2).
\ee
Notice that there is a slight difference between \eqref{SNq0int} and \eqref{AHK}.
In \eqref{SNq0int}, the integrand is symmetric under a permutation of $z_i$ and $z_j$
, while in \eqref{AHK} it is not the case.

Also, Kadell's formula \cite{kad88,kan96} is obtained as a special case of Kaneko's formula \eqref{MacAV}:
\be
\begin{split}
& \int_{[0,1]^N} \de_q z_1 \wedge \de_q z_2 \wedge \dotsm
\wedge \de_q z_N \, P_{\lambda}(z; q, q^k) \prod_{i=1}^N z_i^{u_1} 
\frac{(q z_i; q)_{\infty}}{(q^{u_2+1 } z_i; q)_{\infty}}
\prod_{1 \leq i < j \leq N}
z_i^{2k} \frac{(q^{1-k} z_j/z_i; q)_{\infty}}
{(q^{1+k} z_j/z_i; q)_{\infty}} \cr
&= q^{A_N} 
P_{\lambda}(1,q^k, q^{2k}, \dotsc, q^{(N-1)k}) \cr
& \times \prod_{i=1}^N \frac{\Gamma_q(u_1+1 + (N-i) k+ \lambda_i) \Gamma_q(u_2+1+(N-i)k)
\Gamma_q(1+ik)}
{\Gamma_q(u_1+u_2+2+(2N-i-1)k+\lambda_i ) \Gamma_q(1+k)}.
\end{split}
\ee
Here
\be
P_{\lambda}(1,q^k, q^{2k}, \dotsc, q^{(N-1)k}; q,q^k)
= \frac{(q^{Nk})_{\lambda}^{(q,q^k)}}{h_{\lambda}(q,q^k)}.
\ee
It can be rewritten as follows:
\be
\begin{split}
& \langle P_{\lambda}(z; q,q^k) \rangle \cr
&= \frac{(q^{Nk})_{\lambda}^{(q,q^k)}}{h_{\lambda}(q,q^k)}
\prod_{i=1}^N \frac{\Gamma_q(u_1+1+(N-j) k+\lambda_i)
\Gamma_q(u_1+u_2+2+(2N-j-1)k)}
{\Gamma_q(u_1+1+(N-j)k)\Gamma_q(u_1+u_2+2+(2N-j-1)k+\lambda_i)} \cr
&=\frac{(q^{Nk})_{\lambda}^{(q,q^k)}
(q^{u_1+1+(N-1)k})_{\lambda}^{(q,q^k)}}
{h_{\lambda}(q,q^k) (q^{u_1+u_2+2+2(N-1)k})_{\lambda}^{(q,q^k)}}.
\end{split}
\ee

%%%%%%%%%%%%%%%%%%%%%%%%%%%%%%%%%%%%%%%%%%%%%%%%%%%

%%%%%%%%%%%%%%%%%%%%%%%%%%%%%%%%%%%%%%%%%%


\begin{thebibliography}{99}

\bibitem{LP94}
S.~Lukyanov and Ya.~Pugai, ``Bosonization of ZF algebras: Direction toward
  deformed Virasoro algebra,'' J.\ Exp.\ Theor.\ Phys.\ {\bf 82}, 1021-1045
  (1996) [Zh.\ Eksp.\ Teor.\ Fiz.\ {\bf 109}, 1900-1947 (1996)]
  [arXiv:hep-th/9412128].

\bibitem{FR95}
E.~Frenkel and N.~Reshetikhin, ``Quantum Affine Algebras and Deformations of
  the Virasoso and $\mathcal{W}$-Algebras,'' Commun.\ Math.\ Phys.\ {\bf 178},
  237-264 (1996) [arXiv:q-alg/9505025].

\bibitem{SKAO}
J.~Shiraishi, H.~Kubo, H.~Awata and S.~Odake, ``A quantum deformation of the
  Virasoro algebra and the Macdonald symmetric functions,'' Lett.\ Math.\
  Phys.\ {\bf 38}, 33-51 (1996) [arXiv:q-alg/9507034].

\bibitem{FF95}
B.~Feigin and E.~Frenkel, ``Quantum $\mathcal{W}$-Algebras and Elliptic
  Algebras,'' Commun.\ Math.\ Phys.\ {\bf 178}, 653-678 (1996)
  [arXiv:q-alg/9508009].

\bibitem{AKOS95}
H.~Awata, H.~Kubo, S.~Odake and J.~Shiraishi, ``Quantum $\mathcal{W}_N$
  Algebras and Macdonald Polynomials,'' Commun.\ Math.\ Phys.\ {\bf 179},
  401-416 (1996) [arXiv:q-alg/9508011].

\bibitem{MY95}
K.~Mimachi and Y.~Yamada, ``Singular vectors of the Virasoro Algebra in Terms
  of Jack Symmetric Polynomials,'' Commun.\ Math.\ Phys.\ {\bf 174}, 447-455
  (1995).

\bibitem{mac}
I.~G.~Macdonald, \textit{Symmetric functions and Hall Polynomials}, 2nd ed.,
  Oxford University Press (1995).

\bibitem{AKMOS9604}
H.~Awata, H.~Kubo, Y.~Morita, S.~Odake and J.~Shiraishi, ``Vertex Operators of
  the $q$-Virasoro Algebra; Defining Relations, Adjoint Actions and Four Point
  Functions,'' Lett.\ Math.\ Phys.\ {\bf 41}, 65-78 (1997)
  [arXiv:q-alg/9604023].

\bibitem{kad}
A.~A.~Kadeishvili, ``Vertex operators for deformed Virasoro algebra,'' JETP
  Lett.\ {\bf 63}, 917-923 (1996) [Pisma Zh.\ Eksp.\ Teor.\ Fiz.\ {\bf 63},
  876-881 (1996)] [arXiv:hep-th/9604153].

\bibitem{JS97}
M.~Jimbo and J.~Shiraishi, ``A Coset-Type Construction for the Deformed
  Virasoro Algebra,'' Lett.\ Math.\ Phys.\ {\bf 43}, 173-185 (1998)
  [arXiv:q-alg/9709037].

\bibitem{AGT}
L.~F.~Alday, D.~Gaiotto and Y.~Tachikawa, ``Liouville Correlation Functions
  from Four-dimensional Gauge Theories,'' Lett.\ Math.\ Phys.\ {\bf 9}, 167-197
  (2010) [arXiv:0906.3219 [hep-th]].

\bibitem{Wyl0907}
N.~Wyllard, ``$A_{N-1}$ conformal Toda field theory correlation functions from
  conformal $\mathcal{N}=2$ $SU(N)$ quiver gauge theories,'' JHEP {\bf 0911},
  002 (2009) [arXiv:0907.2189 [hep-th]].

\bibitem{MMM0907}
A.~Marshakov, A.~Mironov and A.~Morozov, ``Combinatorial expansions of
  conformal blocks,'' Theor.\ Math.\ Phys.\ {\bf 164}, 831-852 (2010) [Teor.\
  Mat.\ Fiz.\ {\bf 164}, 3-27 (2010)] [arXiv:0907.3946 [hep-th]]; 

A.~Mironov, S.~Mironov, A.~Morozov and And.~Morozov, ``CFT exercises for the
  needs of AGT,'' arXiv:0908.2064 [hep-th];

A.~Mironov and A.~Morozov, ``The power of Nekrasov functions,'' Phys.\ Lett.\ B
  {\bf 680}, 188-194 (2009) [arXiv:0908.2190 [hep-th]]; 

A.~Mironov and A.~Morozov, ``On AGT relation in the case of $U(3)$,'' Nucl.\
  Phys.\ B {\bf 825}, 1-37 (2010) [arXiv:0908.2569 [hep-th]];

A.~Marshakov, A.~Mironov and A.~Morozov, ``On non-conformal limit of the AGT
  relations,'' Phys.\ Lett.\ B {\bf 682}, 125-129 (2009) [arXiv:0909.2052
  [hep-th]].

\bibitem{DV}
R.~Dijkgraaf and C.~Vafa, ``Toda Theories, Matrix Models, Topological Strings,
  and $N = 2$ Gauge Systems,'' arXiv:0909.2453 [hep-th].

\bibitem{MMM0909b}
A.~Marshakov, A.~Mironov and A.~Morozov, ``Zamolodchikov asymptotic formula and
  instanton expansion in $\mathcal{N}=2$ SUSY $N_f = 2N_c$ QCD,'' JHEP {\bf
  0911}, 048 (2009) [arXiv:0909.3338 [hep-th]]; 

A.~Mironov and A.~Morozov, ``Proving AGT relations in the large-$c$ limit,''
  Phys.\ Lett.\ B {\bf 682}, 118-124 (2009) [arXiv:0909.3531 [hep-th]];

A.~Mironov and A.~Morozov, ``Nekrasov functions and exact Bohr-Zommerfeld
  integrals,'' JHEP {\bf 1004}, 040 (2010) [arXiv:0910.5670 [hep-th]]; 

A.~Mironov and A.~Morozov, ``Nekrasov functions from exact Bohr-Sommerfeld
  periods: the case of $SU(N)$,'' J.\ Phys.\ A {\bf 43}, 195401 (2010)
  [arXiv:0911.2396 [hep-th]].

\bibitem{IMO}
H.~Itoyama, K.~Maruyoshi and T.~Oota, ``The Quiver Matrix Model and 2d-4d
  Conformal Connection,'' Prog.\ Theor.\ Phys.\ {\bf 123}, 957-987 (2010)
  [arXiv:0911.4244 [hep-th]].

\bibitem{MMS0911}
A.~Mironov, A.~Morozov and Sh.~Shakirov, ``Matrix model conjecture for exact BS
  periods and Nekrasov functions,'' JHEP {\bf 1002}, 030 (2010)
  [arXiv:0911.5721 [hep-th]];
  
  A.~Mironov, A.~Morozov and Sh.~Shakirov, ``Conformal blocks as Dotsenko-Fateev
  integral discriminants,'' Int.\ J.\ Mod.\ Phys.\ A {\bf 25}, 3173-3207 (2010)
  [arXiv:1001.0563 [hep-th]];
  
  A.~Mironov, A.~Morozov and And.~Morozov, ``Conformal blocks and generalized
  Selberg integrals,'' Nucl.\ Phys.\ B {\bf 843}, 534-557 (2011)
  [arXiv:1003.5752 [hep-th]]; 

  A.~Mironov, A.~Morozov and Sh.~Shakirov, ``Towards a proof of AGT conjecture by
  methods of matrix models,'' Int.\ J.\ Mod.\ Phys.\ A {\bf 27}, 1230001 (2012)
  [arXiv:1011.5629 [hep-th]];

  A.~Mironov, A.~Morozov and Sh.~Shakirov, ``A direct proof of AGT conjecture at
  $\beta = 1$,'' JHEP {\bf 1102}, 067 (2011) [arXiv:1012.3137 [hep-th]].


\bibitem{FL0912}
V.~A.~Fateev and A.~V.~Litvinov, ``On AGT conjecture,'' JHEP {\bf 1002}, 014
  (2010) [arXiv:0912.0504 [hep-th]];
  
  V.~A.~Fateev and A.~V.~Litvinov, ``Integrable structure, W-symmetry and AGT
  relation,'' JHEP {\bf 1201}, 051 (2012) [arXiv:1109.4042 [hep-th]].


\bibitem{IO5}
H.~Itoyama and T.~Oota, ``Method of generating $q$-expansion coefficients for
  conformal block and $\mathcal{N}=2$ Nekrasov function by $\beta$-deformed
  matrix model,'' Nucl.\ Phys.\ B {\bf 838}, 298-330 (2010) [arXiv:1003.2929
  [hep-th]].

\bibitem{IOYone}
H.~Itoyama, T.~Oota and N.~Yonezawa, ``Massive scaling limit of the
  $\beta$-deformed matrix model of Selberg type,'' Phys.\ Rev.\ D {\bf 82},
  085031 (2010) [arXiv:1008.1861 [hep-th]].

\bibitem{AFLT1012}
V.~A.~Alba, V.~A.~Fateev, A.~V.~Litvinov and G.~M.~Tarnopolskiy, ``On
  Combinatorial Expansion of the Conformal Blocks Arising from AGT
  Conjecture,'' Lett.\ Math.\ Phys.\ {\bf 98}, 33-64 (2011) [arXiv:1012.1312
  [hep-th]].


\bibitem{Wy1109}
N.~Wyllard, ``Coset conformal blocks and $\mathcal{N}=2$ gauge theories,''
  arXiv:1109.4264 [hep-th].

\bibitem{EPSS1110}
B.~Estienne, V.~Pasquier, R.~Santachiara and D.~Serban, ``Conformal blocks in
  Virasoro and W theories: Duality and the Calogero-Sutherland model,'' Nucl.\
  Phys.\ B {\bf 860}, 377-420 (2012) [arXiv:1110.1101 [hep-th]].

\bibitem{ZM1110}
H.~Zhang and Y.~Matsuo, ``Selberg Integral and $SU(N)$ AGT Conjecture,'' JHEP
  {\bf 1112}, 106 (2011) [arXiv:1110.5255 [hep-th]].

\bibitem{BBFLT1111}
A.~A.~Belavin, M.~A.~Bershtein, B.~L.~Feigin, A.~V.~Litvinov and
  G.~M.~Tarnopolsky, ``Instanton moduli spaces and bases in coset conformal
  field theory,'' Commun.\ Math.\ Phys.\ {\bf 319}, 269-301 (2013)
  [arXiv:1111.2803 [hep-th]].

\bibitem{KMZ1207}
S.~Kanno, Y.~Matsuo and H.~Zhang, ``Virasoro constraint for Nekrasov instanton
  partition function,'' JHEP {\bf 1210}, 097 (2012) [arXiv:1207.5658 [hep-th]]; 

  S.~Kanno, Y.~Matsuo and H.~Zhang, ``Extended Conformal Symmetry and Recursion
  Formulae for Nekrasov Partition Function,'' JHEP {\bf 1308}, 028 (2013)
  [arXiv:1306.1523 [hep-th]].

\bibitem{tan1301}
M.-C.~Tan,
  ``M-theoretic derivations of 4d-2d dualities: from a geometric Langlands duality for surfaces, to the AGT  
   correspondence, to integrable systems,''
  JHEP {\bf 1307}, 171 (2013)
  [arXiv:1301.1977 [hep-th]].
  
\bibitem{MS1307}
A.~Morozov and A.~Smirnov, ``Towards the Proof of AGT Relations with the Help
  of the Generalized Jack Polynomials,'' Lett.\ Math.\ Phys.\ {\bf 104} 585-612
  (2014) [arXiv:1307.2576 [hep-th]].

\bibitem{MMZ1312}
S.~Mironov, And.~Morozov and Y.~Zenkevich, ``Generalized Jack polynomials and
  the AGT relations for the $SU(3)$ group,'' JETP Lett.\ {\bf 99}, 109-113
  (2014) [arXiv:1312.5732 [hep-th]].

\bibitem{MRZ1405}
Y.~Matsuo, C.~Rim and H.~Zhang, ``Construction of Gaiotto states with
  fundamental multiplets through degenerate DAHA,'' JHEP {\bf 1409}, 028 (2014)
  [arXiv:1405.3141 [hep-th]].

\bibitem{IY1507}
H.~Itoyama and R.~Yoshioka, ``Developments of theory of effective prepotential
  from extended Seiberg-Witten system and matrix models,'' Prog.\ Theor.\ Exp.\
  Phys.\ 11B103 (2015) [arXiv:1507.00260 [hep-th]].

\bibitem{NZ1511}
A.~Nedelin and M.~Zabzine,
  ``q-Virasoro constraints in matrix models,''
  arXiv:1511.03471 [hep-th].

\bibitem{AY0910}
H.~Awata and Y.~Yamada, ``Five-dimensional AGT conjecture and the deformed
  Virasoro algebra,'' JHEP {\bf 1001}, 125 (2010) [arXiv:0910.4431 [hep-th]].

\bibitem{AY1004}
H.~Awata and Y.~Yamada, ``Five-Dimensional AGT Relation and the Deformed
  $\beta$-Ensemble,'' Prog.\ Theor.\ Phys.\ {\bf 124}, 227-262 (2010)
  [arXiv:1004.5122 [hep-th]].

\bibitem{MMSS1105}
A.~Mironov, A.~Morozov, Sh.~Shakirov and A.~Smirnov, ``Proving AGT conjecture
  as HS duality: Extension to five dimensions,'' Nucl.\ Phys.\ B {\bf 855},
  128-151 (2012) [arXiv:1105.0948 [hep-th]].

\bibitem{NPP1303}
F.~Nieri, S.~Pasquetti and F.~Passerini, ``3d and 5d Gauge Theory Partition
  Functions as $q$-deformed CFT Correlators,'' Lett.\ Math.\ Phys.\ {\bf 105},
  109-148 (2015) [arXiv:1303.2626 [hep-th]].

\bibitem{Tan1309}
M.-C.~Tan, ``An M-theoretic derivation of a 5d and 6d AGT correspondence, and
  relativistic and elliptized integrable systems,'' JHEP {\bf 1312}, 031 (2013)
  [arXiv:1309.4775 [hep-th]].

\bibitem{NPPT1312}
F.~Nieri, S.~Pasquetti, F.~Passerini and A.~Torrielli, ``5D partition
  functions, $q$-Virasoro systems and integrable spin-chains'' JHEP {\bf 1412},
  040 (2014) [arXiv:1312.1294 [hep-th]].

\bibitem{ohk1404}
Y.~Ohkubo, ``Existence and orthogonality of generalized Jack polynomials and
  its $q$-deformation,'' arXiv:1404.5401 [math-ph].

\bibitem{zen1412}
Y.~Zenkevich, ``Generalized Macdonald polynomials, spectral duality for
  conformal blocks and AGT correspondence in five dimensions,'' JHEP {\bf
  1505}, 131 (2015) [arXiv:1412.8592 [hep-th]].

\bibitem{AFO1512}
H.~Awata, H.~Fujino and Y.~Ohkubo, ``Crystallization of deformed Virasoro
  algebra, Ding-Iohara-Miki algebra and 5D AGT correspondence,''
  arXiv:1512.08016 [math-ph].

\bibitem{KP1512}
T.~Kimura and V.~Pestun, ``Quiver W-algebras,'' arXiv:1512.08533 [hep-th].

\bibitem{MZ1510}
A.~Morozov and Y.~Zenkevich, ``Decomposing Nekrasov Decomposition,''
  arXiv:1510.01896 [hep-th].

\bibitem{IKY1511}
A.~Iqbal, C.~Koz\c{c}az and S.-T.~Yau, ``Elliptic Virasoro Conformal Blocks,''
  arXiv:1511.00458 [hep-th].

\bibitem{nie1511}
F.~Nieri, ``An elliptic Virasoro symmetry in 6d,'' arXiv:1511.00574 [hep-th].

\bibitem{MMZ1512}
A.~Mironov, A.~Morozov and Y.~Zenkevich, ``On elementary proof of AGT relations
  from six dimensions,'' arXiv:1512.06701 [hep-th].

\bibitem{kan96}
J.~Kaneko, ``$q$-Selberg integrals and Macdonald polynomials,'' Ann.\ scient.\
  \'{E}c.\ Norm.\ Sup.\ {\bf 29}, 583-637 (1996).

\bibitem{BF1105}
V.~Belavin and B.~Feigin, ``Super Liouville conformal blocks from
  $\mathcal{N}=2$ $SU(2)$ quiver gauge theories,'' JHEP {\bf 1107}, 079 (2011)
  [arXiv:1105.5800 [hep-th]].

\bibitem{NT1106}
T.~Nishioka and Y.~Tachikawa, ``Central charges of para-Liouville and Toda
  theories from M-5-branes,'' Phys.\ Rev.\ D {\bf 84}, 046009 (2011)
  [arXiv:1106.1172 [hep-th]].

\bibitem{BBB1106}
A.~Belavin, V.~Belavin and M.~Bershtein, ``Instantons and 2d Superconformal
  field theory,'' JHEP {\bf 1109}, 117 (2011) [arXiv:1106.4001 [hep-th]].

\bibitem{BMT}
G.~Bonelli, K.~Maruyoshi and A.~Tanzini, ``Instantons on ALE spaces and super
  Liouville conformal field theories,'' JHEP {\bf 1108}, 056 (2011)
  [arXiv:1106.2505 [hep-th]]; 
  
  G.~Bonelli, K.~Maruyoshi and A.~Tanzini, 
  ``Gauge Theories on ALE Space and Super Liouville
  Correlation Functions,'' Lett.\ Math.\ Phys.\ {\bf 101}, 103-124 (2012)
  [arXiv:1107.4609 [hep-th]].

\bibitem{I1110}
Y.~Ito, ``Ramond sector of super Liouville theory from instantons on an ALE
  space,'' Nucl.\ Phys.\ B {\bf 861}, 387-402 (2012) [arXiv:1110.2176
  [hep-th]].

\bibitem{AT1110}
M.~N.~Alfimov and G.~M.~Tarnopolsky, ``Parafermionic Liouville field theory and
  instantons on ALE spaces,'' JHEP {\bf 1202}, 036 (2012) [arXiv:1110.5628
  [hep-th]].

\bibitem{BW1205}
V.~Belavin and N.~Wyllard, ``$\mathcal{N}=2$ superconformal blocks and
  instanton partition functions,'' JHEP {\bf 1206}, 173 (2012) [arXiv:1205.3091
  [hep-th]].

\bibitem{BM1210}
A.~Belavin and B.~Mukhametzhanov, ``$N=1$ superconformal blocks with Ramond
  fields from AGT correspondence,'' JHEP {\bf 1301}, 178 (2013)
  [arXiv:1210.7454 [hep-th]].

\bibitem{BBT1211}
A.~A.~Belavin, M.~A.~Bershtein and G.~M.~Tarnopolsky, ``Bases in coset
  conformal field theory from AGT correspondence and Macdonald polynomials at
  the roots of unity,'' arXiv:1211.2788 [hep-th].

\bibitem{ABT1306}
M.~N.~Alfimov, A.~A.~Belavin and G.~M.~Tarnopolsky, ``Coset conformal field
  theory and instanton counting on $\mathbb{C}^2/\mathbb{Z}_p$,'' JHEP {\bf
  1308}, 134 (2013) [arXiv:1306.3938 [hep-th]].

\bibitem{IOY2}
H.~Itoyama, T.~Oota and R.~Yoshioka, ``2d-4d Connection between $q$-Virasoro/W
  Block at Root of Unity Limit and Instanton Partition Function on ALE Space,''
  Nucl.\ Phys.\ B {\bf 877}, 506-537 (2013) [arXiv:1308.2068 [hep-th]];
  
  H.~Itoyama, T.~Oota and R.~Yoshioka,
  ``q-Virasoro algebra at root of unity limit and 2d-4d connection,'' 
  J.\ Phys.\ Conf.\ Ser.\ {\bf 474}, 012022 (2013);

  H.~Itoyama, T.~Oota and R.~Yoshioka, ``$q$-Virasoro/W Algebra at Root of Unity
  and Parafermions,'' Nucl.\ Phys.\ B {\bf 889}, 25-35 (2014) [arXiv:1408.4216
  [hep-th]].

\bibitem{spo1409}
L.~Spodyneiko,
  ``AGT correspondence: Ding-Iohara algebra at roots of unity and Lepowsky–Wilson construction,''
  J.\ Phys.\ A {\bf 48}, 275404 (2015)
  [arXiv:1409.3465 [hep-th]].

\bibitem{yos1512}
R.~Yoshioka, ``The integral representation of solutions of KZ equation and a
  modification by $\mathcal{K}$ operator insertion,'' arXiv:1512.01084
  [hep-th].

\bibitem{nek0206}
N.~A.~Nekrasov, ``Seiberg-Witten Prepotential From Instanton Counting,'' Adv.\
  Theor.\ Math.\ Phys.\ {\bf 7}, 831-864 (2004) [arXiv:hep-th/0206161].

\bibitem{NO0306}
N.~Nekrasov and A.~Okounkov, ``Seiberg-Witten Theory and Random Partitions,''
  in \textit{The Unity of Mathematics}, in Honor of the Ninetieth Birthday of
  I.M. Gelfand, Progress of Mathematics Vol. \textbf{244}, 525-596, ed. by P.
  Etingof, V. Retakh and I.M. Singer, Birkh''{a}user, Boston (2006)
  [arXiv:hep-th/0306238].

\bibitem{NY0505}
H. Nakajima and K. Yoshioka, ``Instanton counting on blowup. II. $K$-theoretic
  partition function,'' Transfor. Groups \textbf{10}, 489-519 (2005)
  [arXiv:math/0505553 [math.AG]].

\bibitem{aom98}
K.~Aomoto, ``On Elliptic Product Formulas for Jackson Integrals Associated with
  Reduced Root Systems,'' J.\ Alg.\ Geom.\ {\bf 8}, 115-126 (1998).

\bibitem{ask}
R.~Askey, ``Some Basic Hypergeometric Extensions of Integrals of Selberg and
  Andrews,'' SIAM J.\ Math.\ Anal.\ {\bf 11}, 938-951 (1980).

\bibitem{hab}
L.~Habsieger, ``Une $q$-Int\'{e}grale de Selberg et Askey,'' SIAM J.\ Math.\
  Anal.\ {\bf 19}, 1475-1489 (1988).

\bibitem{kad88}
K.~W.~J.~Kadell, ``A Proof of Askey's Conjectured $q$-Analogue of Selberg's
  Integral and a Conjecture of Morris,'' SIAM J.\ Math.\ Anal.\ {\bf 19},
  969-986 (1988).

  
  
\end{thebibliography}
\end{document}